\documentclass{moriond}

\def\Journal#1#2#3#4{{#1} {\bf #2}, #3 (#4)}

\def\PRL{\em Phys. Rev. Lett.}

\def\JCAP{{\em JCAP}}
\def\PTEP{{\em PTEP}}
\def\NRP{{\em Nature Rev. Phys.}}

\def\be{\begin{equation}}
\def\ee{\end{equation}}
\def\bea{\begin{eqnarray}}
\def\eea{\end{eqnarray}}

\usepackage{amsmath} 

\newcommand{\Photo}{\includegraphics[height=35mm]{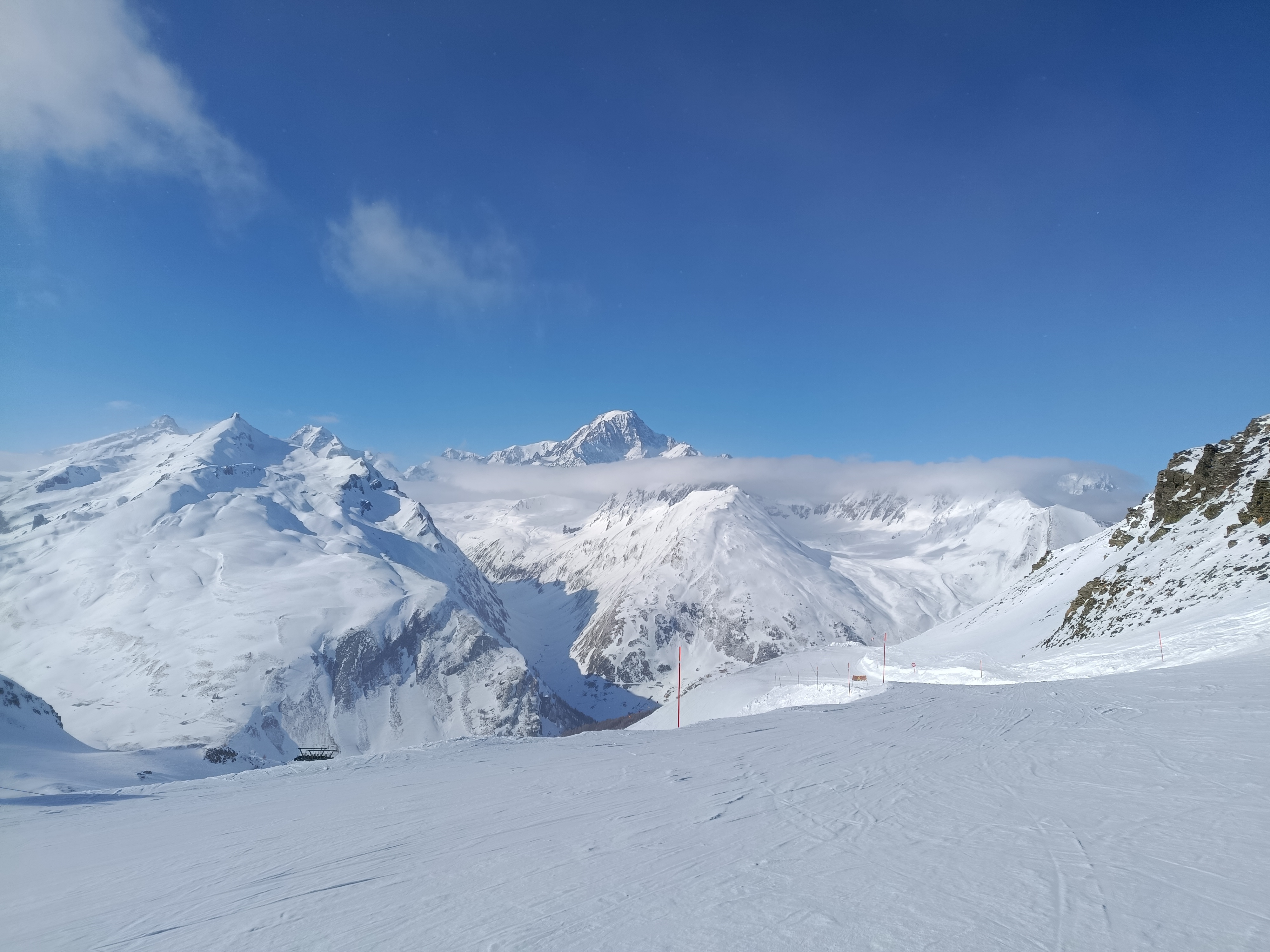}}

\begin{document}
\vspace*{4cm}
\title{Field-level constraints on cosmic birefringence with a hybrid $\boldsymbol{E}$--$\boldsymbol{B}$ internal linear combination}

\author{ Mathieu Remazeilles }

\address{Instituto de Física de Cantabria (CSIC-UC),\\
Avda. de los Castros s/n, 39005 Santander, Spain}

\maketitle

\abstracts{
Cosmic birefringence, a signature of parity-violating physics, rotates the cosmic microwave background (CMB) polarization plane, generating correlations between CMB $E$- and $B$-mode anisotropies. Measuring this effect remains challenging due to degeneracies with spurious rotations from instrumental polarization angle miscalibration and limited knowledge of Galactic foreground $EB$ correlations. We present a blind, map-based approach based on a multi-Stokes \emph{hybrid} internal linear combination (ILC) that breaks this degeneracy and disentangles correlated and uncorrelated CMB polarization components. By jointly combining $E$- and $B$-mode frequency maps, the method preserves achromatic birefringence-induced CMB anisotropies while downweighting foregrounds and chromatic CMB anisotropies resulting from instrumental miscalibration. This enables a direct spatial linear-regression estimator of the birefringence angle. Applied to \emph{LiteBIRD} simulations, the method yields competitive constraints on birefringence. Applied to \emph{Planck} PR4 data, we measure a birefringence angle $\beta \simeq 0.32^\circ \pm 0.12^\circ$, consistent with previous independent analyses and stable across sky fractions.
 }

\section{Introduction}

Parity-violating physics beyond the Standard Model allows dynamical pseudo-scalar fields such as axion-like particles to couple to CMB photons via a Chern--Simons term. In this case, the Universe behaves as a \emph{birefringent} medium which rotates the CMB polarization by a small angle $\beta$, thus mixing $E$- and $B$-modes \cite{rev}
 \begin{align}
 \label{eq:rotation}
\begin{cases}
E_{\ell m}^{\,\rm CMB,\, rot.}(\nu) \simeq E_{\ell m}^{\,\rm CMB} - 2\beta B_{\ell m}^{\,\rm CMB} -  2\alpha(\nu) B_{\ell m}^{\,\rm CMB}\\
B_{\ell m}^{\,\rm CMB,\, rot.}(\nu) \simeq B_{\ell m}^{\,\rm CMB} + 2\beta E_{\ell m}^{\,\rm CMB} +  2\alpha(\nu) E_{\ell m}^{\,\rm CMB}
\end{cases}\,,
\end{align}
and inducing a parity-violating $EB$ correlation between CMB $E$- and $B$-mode anisotropies:
\begin{align}
 \label{eq:eb}	
C_\ell^{EB,\,{\rm CMB,\, rot.}}(\nu) = 2\left(\beta + \alpha(\nu)\right)\left(C_\ell^{EE,{\rm CMB}} - C_\ell^{BB,{\rm CMB}}\right)\,.
\end{align}

Measuring $\beta$ via the $EB$ cross-power spectrum enables, in principle, constraints on either the dynamics of the axion field or its coupling to photons. However, polarization angle miscalibration in the instrument detectors also rotates the observed CMB polarization by a random and chromatic (i.e., frequency-dependent) angle $\alpha(\nu)$. This induces a degeneracy between $\beta$ and $\alpha$ in the observed $EB$ cross-spectrum (Eq.~\ref{eq:eb}), making their separation the main challenge.

An interesting approach was proposed to break this degeneracy \cite{mk} by comparing the observed $EB$ spectrum in CMB-dominated and Galactic foreground-dominated sky regions and frequency ranges, since Galactic emission is unaffected by cosmic birefringence and therefore probes only $\alpha$. This yielded a detection of a non-zero birefringence angle, $\beta = 0.30^\circ \pm 0.11^\circ$, from \emph{Planck} PR4 data \cite{pr4}. However, this fully-parametric approach relies on modelling assumptions for Galactic $EB$ correlations, which remain uncertain, and the inferred constraints on $\beta$ were found to vary with sky fraction \cite{pr4}, which may challenge the interpretation of the signal as purely cosmological.%which may suggest residual foreground or instrumental systematic effects.

In this work \cite{ebilc}, we propose a blind (i.e. without explicit foreground modelling), map-based ILC method to break the $\alpha$--$\beta$ degeneracy. The key observation is that the cosmological rotation $\beta$ is \emph{achromatic}, whereas the instrumental rotation $\alpha$ is \emph{chromatic}, and ILC weights respond differently to achromatic and chromatic effects. We further develop a \emph{hybrid} $E$--$B$ ILC that jointly combines $E$- and $B$-mode frequency maps to disentangle the correlated and uncorrelated components of the CMB polarization field. Overall, our approach preserves the achromatic birefringence-induced CMB anisotropies while downweighting chromatic contributions from instrumental miscalibration and foreground contamination, and suppresses the uncorrelated part of CMB $E$-mode anisotropies, thereby enabling a direct spatial linear regression to estimate $\beta$.

\section{Methodology}

\subsection{An ILC to break degeneracy between achromatic birefringence and chromatic miscalibration}

The ILC constructs a cleaned CMB map as a weighted sum of multi-frequency observations. The weights $w(\nu)$ are constrained to sum to unity, $\sum_\nu w(\nu) = 1$, ensuring preservation of achromatic signals such as the CMB.

This property has important implications for polarization rotation. The cosmic birefringence angle $\beta$ is achromatic and therefore fully preserved by the ILC. In contrast, the instrumental miscalibration angle $\alpha(\nu)$ is chromatic and therefore modulated by the ILC weights. As the weights vary with multipole and sky position, the projected miscalibration angle  ${\bar{\alpha}_\ell = \sum_\nu w_\ell(\nu) \alpha(\nu)}$ becomes scale-dependent and spatially varying in addition to being downweighted.

\begin{figure}[t]
\centering
\includegraphics[width=0.38\textwidth]{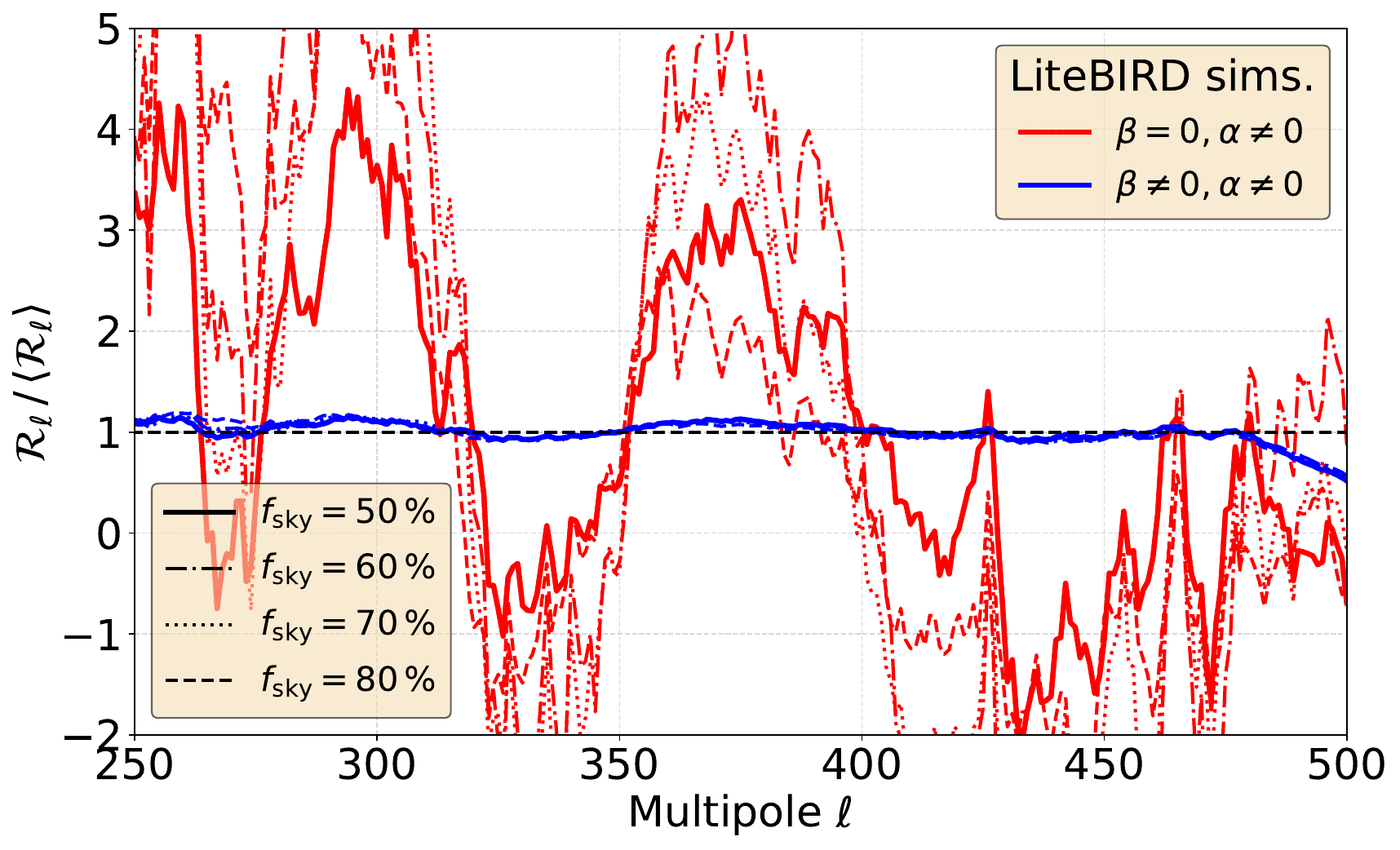}~
\includegraphics[width=0.38\textwidth]{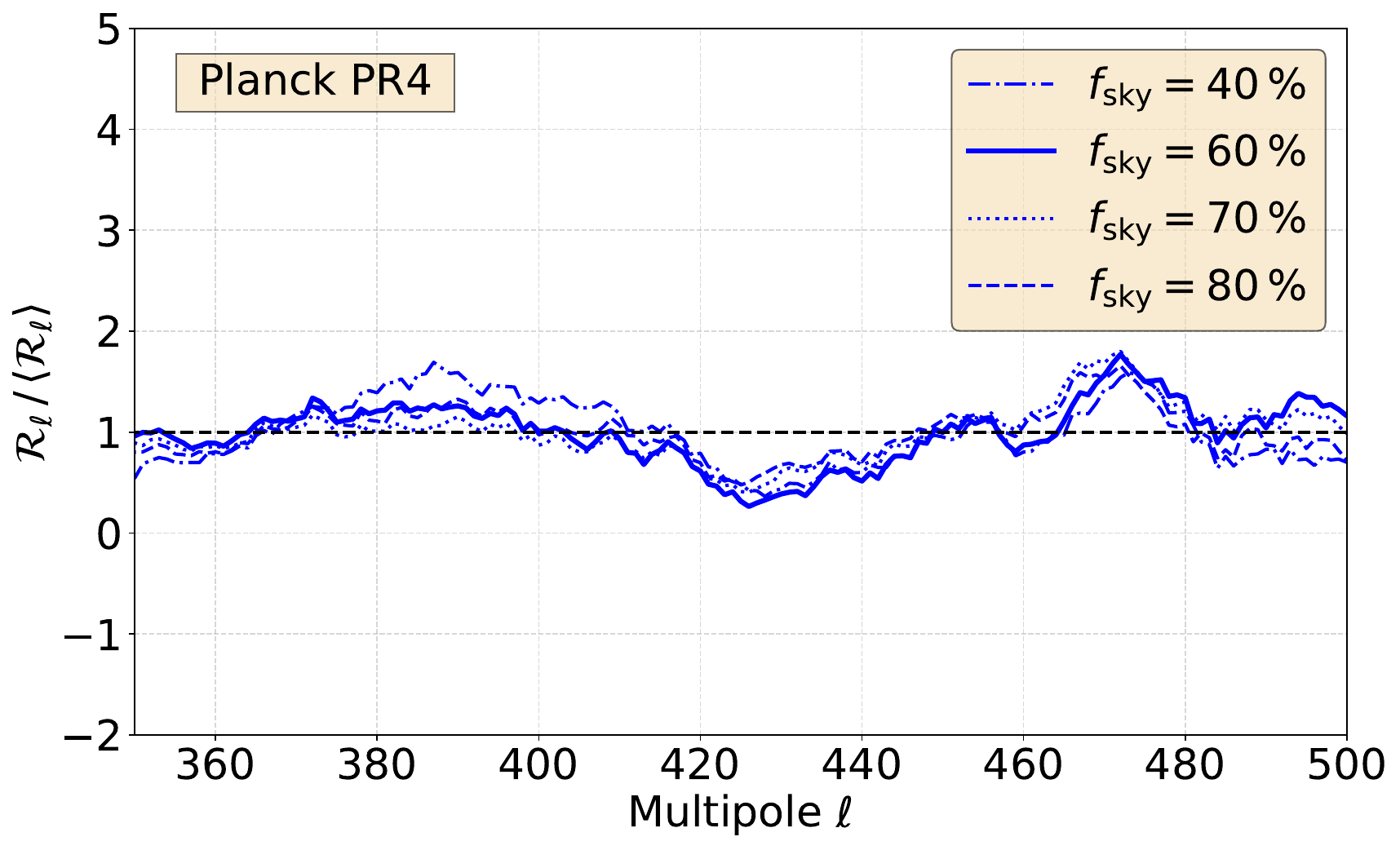}
\caption{Diagnostic ratio $R_\ell$ (Eq.~\ref{eq:ratio}), normalized by its mean, for \emph{LiteBIRD} simulations (left) and \emph{Planck} PR4 data (right). In the absence of cosmic birefringence (red lines), $R_\ell$ varies with multipole and sky fraction due to spatial modulation of the chromatic miscalibration angles by the ILC weights. In the presence of achromatic cosmic birefringence (blue lines), it becomes relatively flat and constant across multipoles and sky fractions.}
\label{fig:Rell}
\end{figure}

This distinction provides a way to disentangle cosmological and instrumental rotation effects. The $EB$ cross-spectrum between the ILC CMB $E$- and $B$-mode maps differs from Eq.~(\ref{eq:eb}) as 
\begin{equation}
\label{eq:ebilc}
C_\ell^{EB,\,{\rm ILC}} = 2\left(\beta + \bar{\alpha}_\ell^{(B)}\right)C_\ell^{EE,\,{\rm CMB}} - 2\left(\beta + \bar{\alpha}_\ell^{(E)}\right)C_\ell^{BB,\,{\rm CMB}}\,, 
\end{equation}
where the projected miscalibration angles $\bar{\alpha}_\ell^{(E)}$, $\bar{\alpha}_\ell^{(B)}$ inherit their $\ell$-dependence from $E$- and $B$-mode ILC weights. As a result, we can build a powerful observable for birefringence diagnostics:
\begin{equation}
\label{eq:ratio}
R_\ell = \frac{C_\ell^{EB,\,{\rm ILC}}}{2\left(C_\ell^{EE} - C_\ell^{BB}\right)}\simeq 
\begin{cases}
f(\ell)\quad {\rm if}\, \beta=0\\
\beta\, (\sim{\rm constant})\quad {\rm if}\, \beta\neq 0 \gg \bar{\alpha}
\end{cases}\,,
\end{equation}
where $C_\ell^{EE}$ and $C_\ell^{BB}$ are $\Lambda{\rm CDM}$ spectra. In the absence of birefringence, $R_\ell$ varies with multipole and sky fraction due to the projected miscalibration angles. In contrast, when birefringence dominates ($\beta \gg \bar{\alpha}_\ell$), $R_\ell$ becomes nearly flat and constant, with little variations across multipoles and across the sky. This behaviour is illustrated in Figure~\ref{fig:Rell} for both \emph{LiteBIRD} simulations and \emph{Planck} PR4 data, where the flatness of $R_\ell$ provides a clear signature of cosmic birefringence.

\subsection{A hybrid $E$--$B$ ILC to disentangle correlated and uncorrelated CMB $E$-mode anisotropies}

Standard ILC methods treat $E$- and $B$-mode frequency maps independently and preserve the full CMB signal, as long as it is uncorrelated with foregrounds. However, birefringence induces correlations between CMB $E$- and $B$-modes.

We introduce a multi-Stokes \emph{hybrid} ILC combining $E$- and $B$-mode frequency maps while targeting the CMB $E$-mode signal. In this framework, CMB $B$-modes act as correlated contaminants. As a result, in the presence of birefringence, the CMB $E$-mode component correlated with $B$-modes is suppressed through variance minimization.

Therefore, the Hybrid $E$--$B$ ILC reconstructs the \emph{uncorrelated} component of CMB $E$-mode anisotropies, i.e. $\langle \hat{E}^{\,\rm Hybrid\, ILC}, B\rangle = 0$, regardless of whether cosmic birefringence is present. Conversely, the difference between the standard ILC and Hybrid ILC CMB $E$-mode maps isolates the  \emph{correlated} component of CMB $E$-mode anisotropies induced by cosmic birefringence.

\subsection{From simple correlation to linear regression}

The total CMB $E$-mode field, $\tilde{E}^{\,\rm ILC}$, as obtained from the standard ILC map, can be decomposed into an \emph{uncorrelated} part, $\hat{E}^{\,\rm hybrid\, ILC}$, as reconstructed by the Hybrid $E$--$B$ ILC, and a \emph{correlated} part proportional to the CMB $B$-mode field, $\tilde{B}^{\,\rm ILC}$ also obtained from the standard ILC:
\begin{equation}
\label{eq:split}
\tilde{E}^{\,\rm ILC} = \hat{E}^{\,\rm Hybrid\, ILC} + \frac{\langle\tilde{E}^{\,\rm ILC},\tilde{B}^{\,\rm ILC}\rangle}{\langle\tilde{B}^{\,\rm ILC},\tilde{B}^{\,\rm ILC}\rangle}\tilde{B}^{\,\rm ILC}\,,
\end{equation}
as $\langle\hat{E}^{\,\rm Hybrid\, ILC},\tilde{B}^{\,\rm ILC}\rangle = 0$ by construction. 

According to the expression of the $EB$ cross-spectrum after ILC projection (Eq.~\ref{eq:ebilc}), the correlation coefficient $\langle\tilde{E}^{\,\rm ILC},\tilde{B}^{\,\rm ILC}\rangle$ in front of the $B$-mode field in Eq.~\eqref{eq:split} can be split into a contribution proportional to the constant birefringence angle $\beta$ and another contribution proportional to the variable, scale-dependent projected miscalibration angles $\bar{\alpha}^{(E)}_\ell$, $\bar{\alpha}^{(B)}_\ell$.

Therefore, the difference between standard and Hybrid ILC $E$-mode maps linearly regresses with $B$-modes:
\begin{equation}
\label{eq:regression}
\tilde{E}^{\,\rm ILC}(\hat{n}) - \hat{E}^{\,\rm Hybrid\, ILC}(\hat{n}) = \beta \, (F \ast \tilde{B}^{\,\rm ILC})(\hat{n}) + \varepsilon(\hat{n})\,,
\end{equation}
with the birefringence angle $\beta$ as the constant regression coefficient, and $F$ a modulation kernel, which only depends on $\Lambda{\rm CDM}$ $EE$ and $BB$ spectra in harmonic space and convolves the $\tilde{B}^{\,\rm ILC}$ field in map space.
Importantly, contributions from chromatic miscalibration angles $\alpha$ and foregrounds are downweighted and absorbed into a spatially varying noise term $\varepsilon(\hat{n})$ due to their modulation by ILC weights, thus preventing bias in the spatial regression.

\begin{figure}[t]
\centering
\begin{minipage}{0.6\linewidth}
\hspace{-0.2cm}
\includegraphics[width=0.3\textwidth]{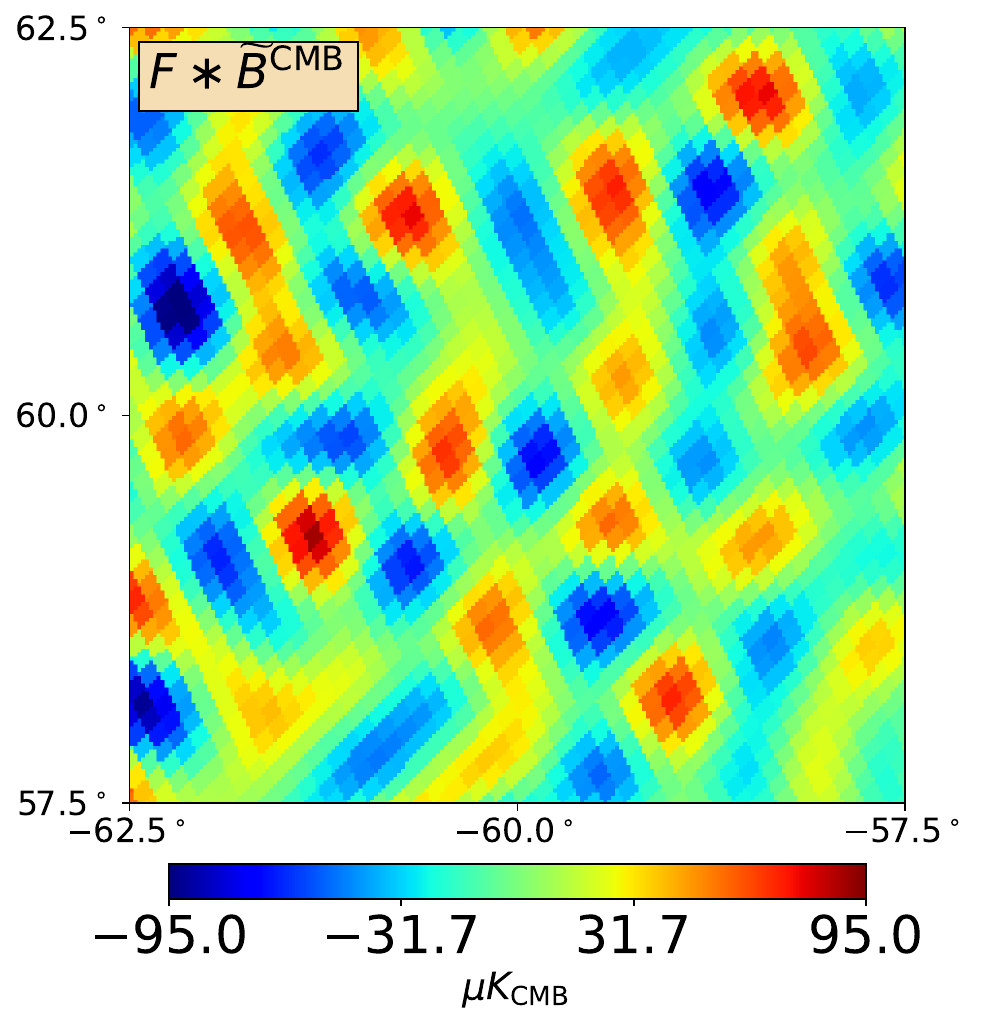}~
\hspace{-0.5cm}
\includegraphics[width=0.3\textwidth]{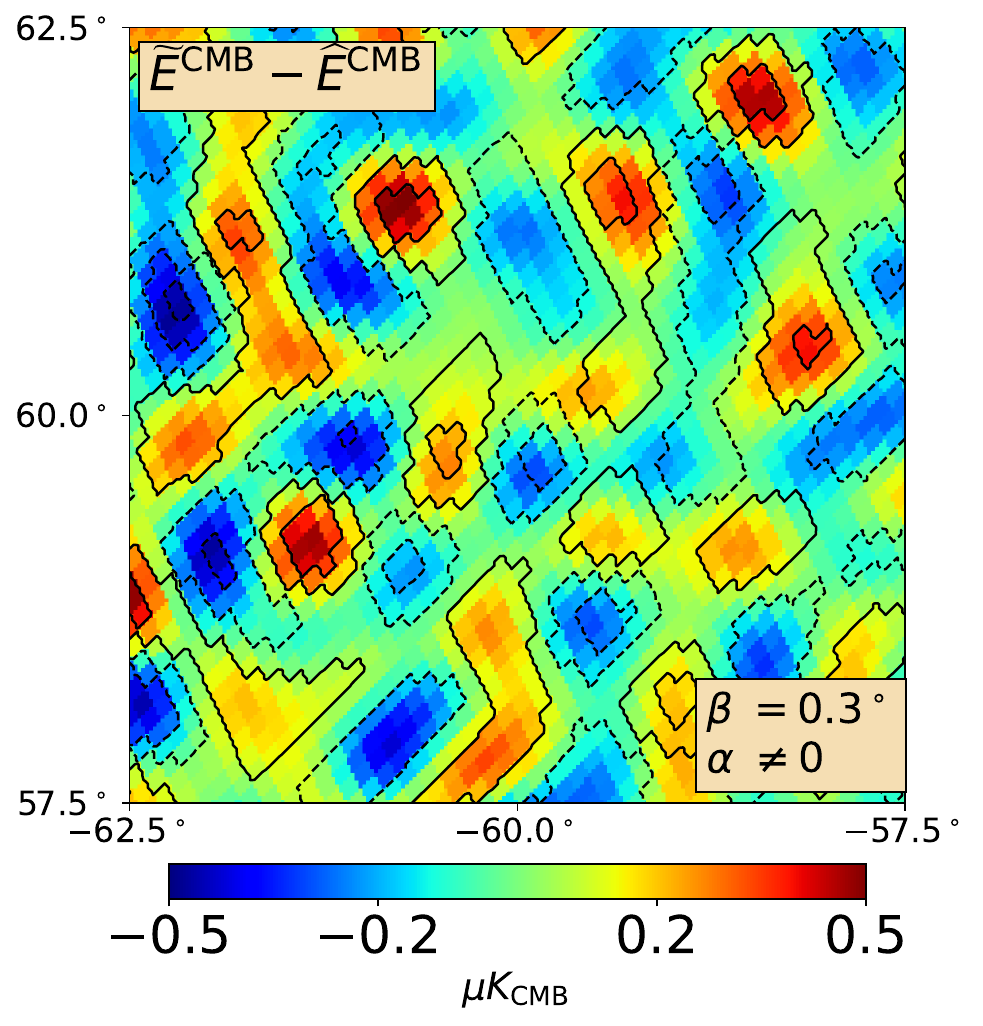}~
\hspace{-0.5cm}
\includegraphics[width=0.3\textwidth]{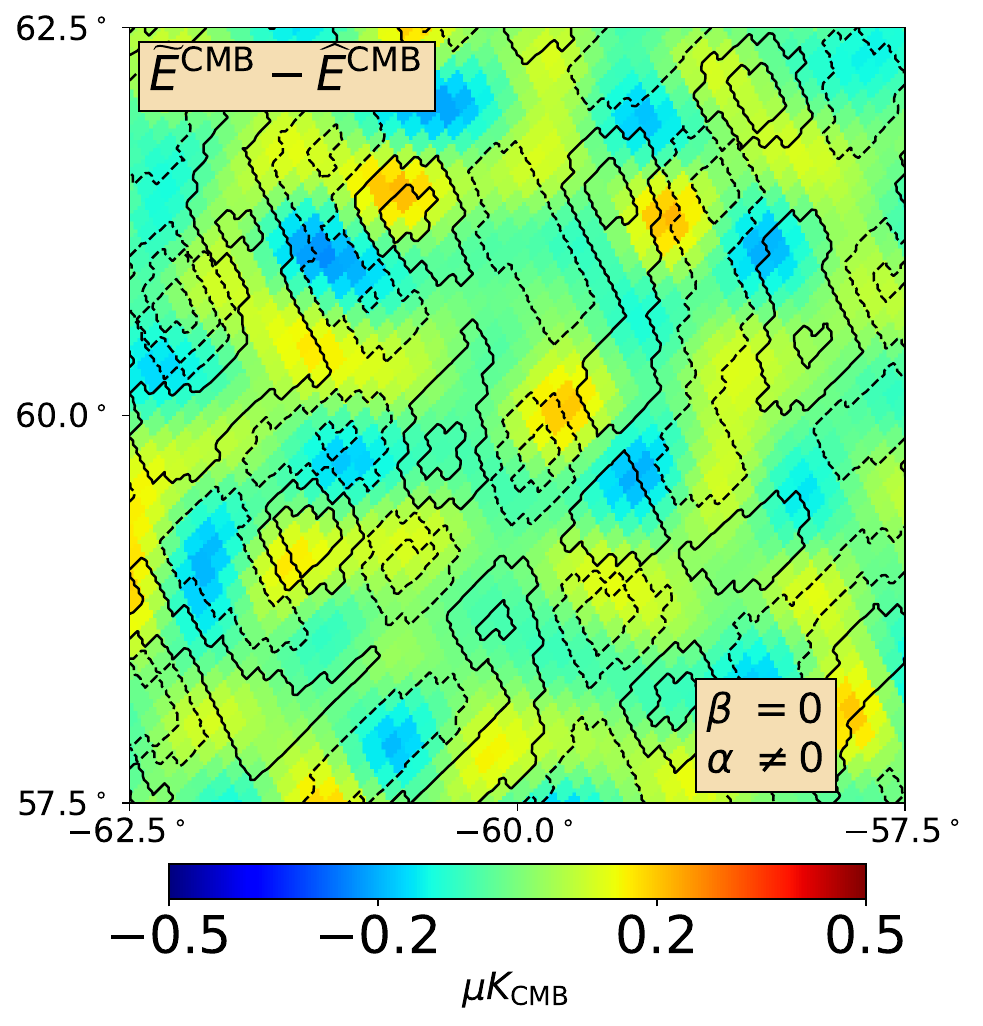}
\end{minipage}~
\hfill
\hspace{-1.4cm}
\begin{minipage}{0.59\linewidth}
\includegraphics[width=0.25\textwidth]{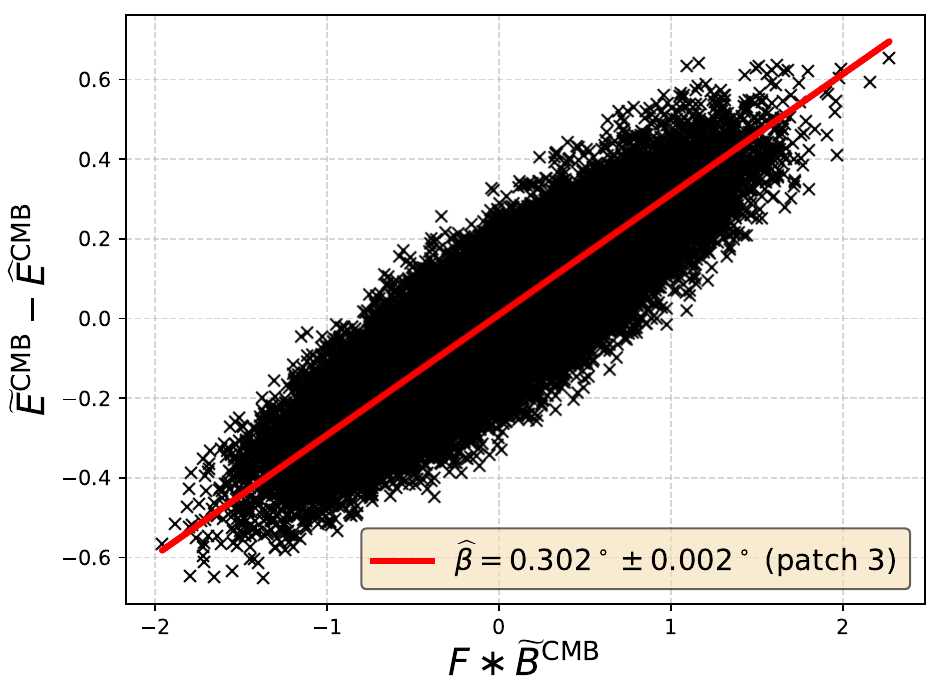}~
\includegraphics[width=0.25\textwidth]{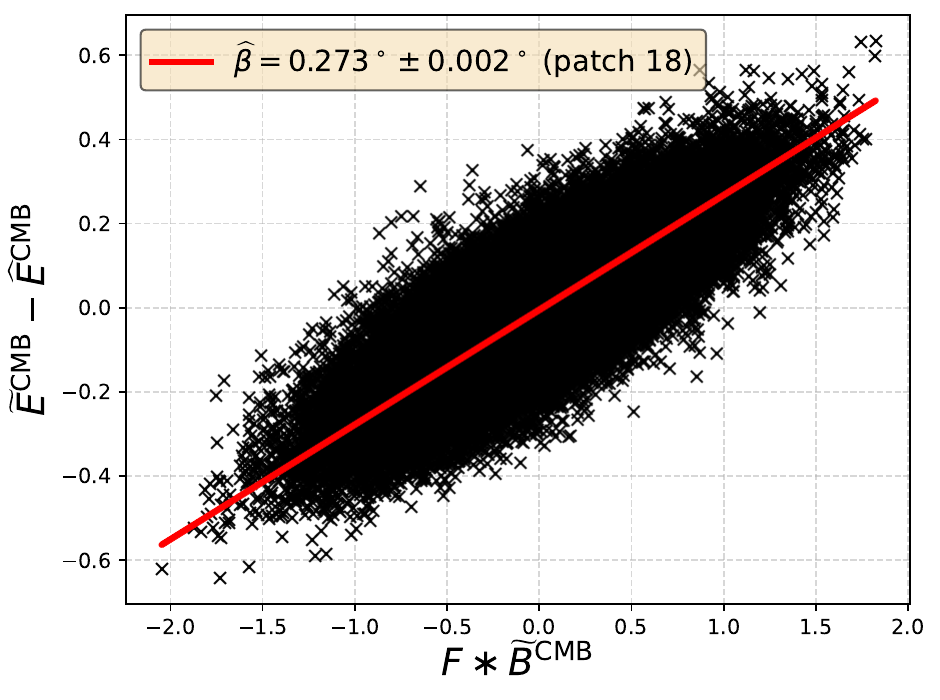}~
\includegraphics[width=0.25\textwidth]{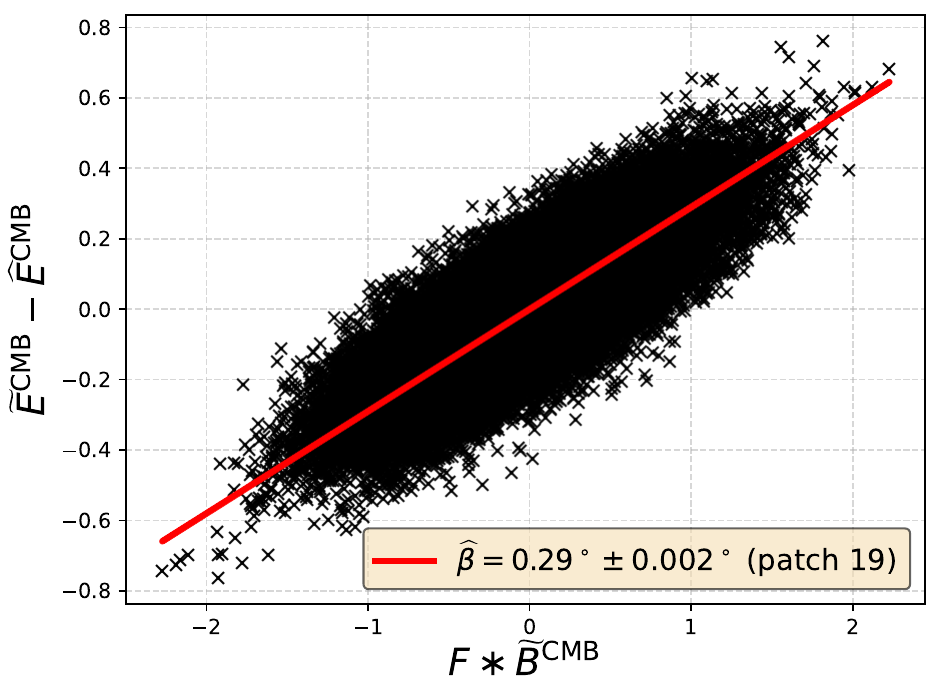}\\
\includegraphics[width=0.25\textwidth]{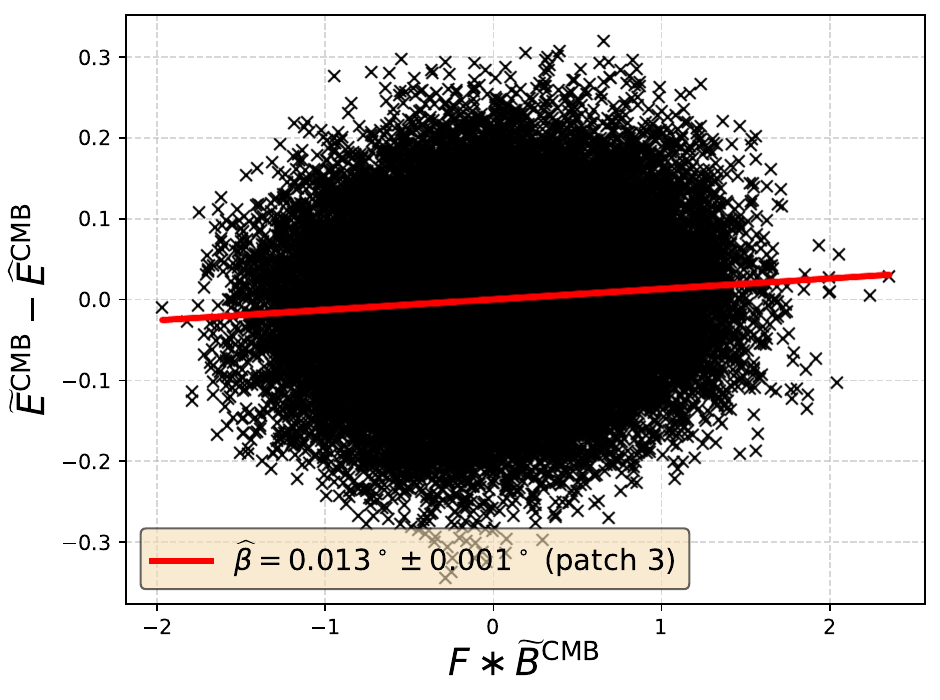}~
\includegraphics[width=0.25\textwidth]{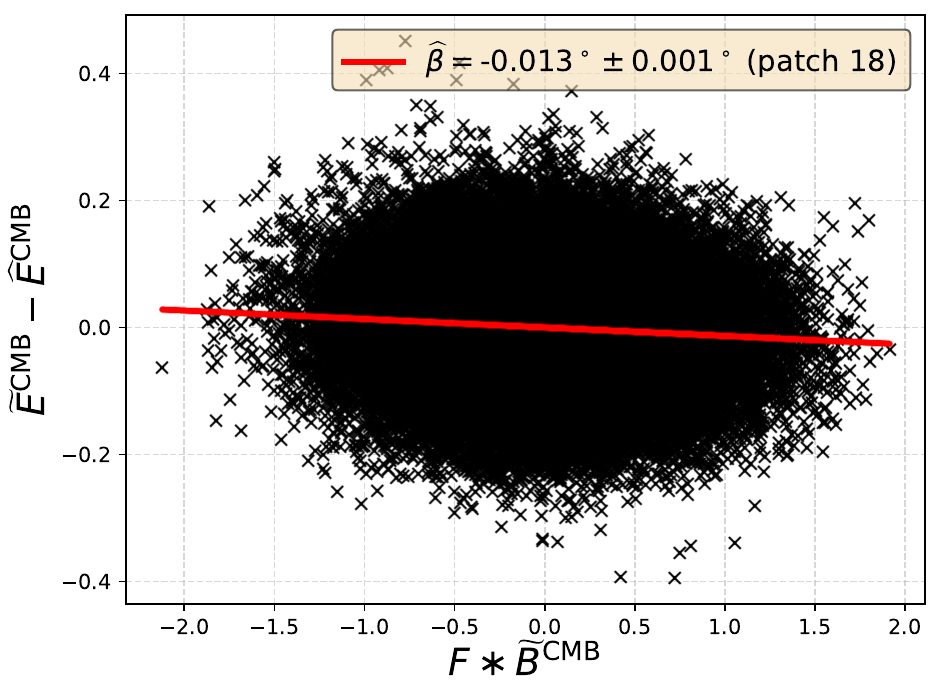}~
\includegraphics[width=0.25\textwidth]{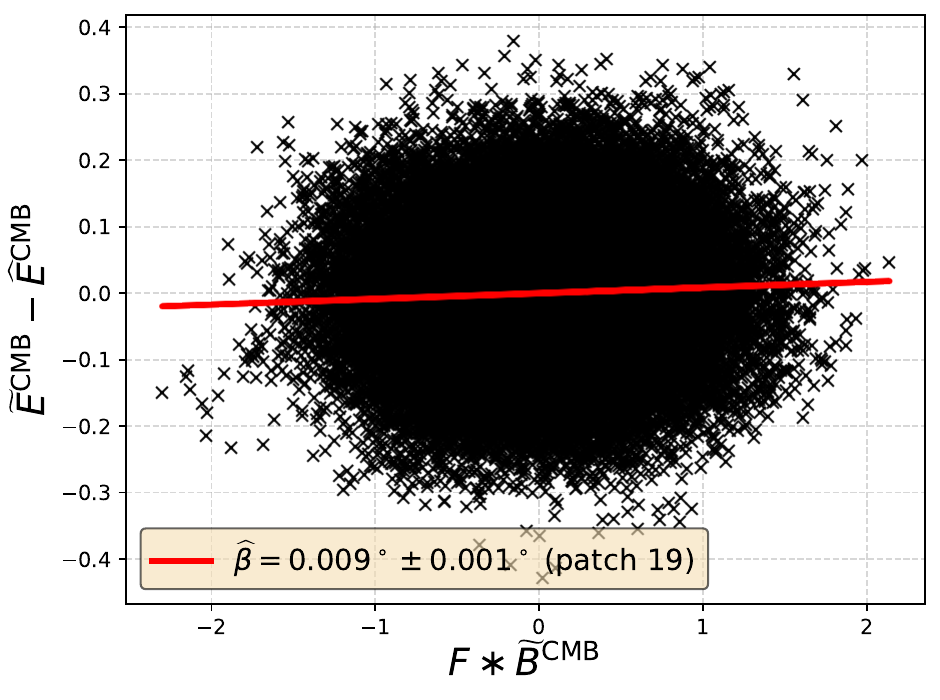}
\end{minipage}
\caption{Regression method on \emph{LiteBIRD} simulations. \emph{Left block}: $F\ast\tilde{B}^{\,\rm ILC}$ field (left), and difference between standard and Hybrid ILC maps, $\tilde{E}^{\,\rm ILC} - \hat{E}^{\,\rm Hybrid, ILC}$, for $\beta\neq0$ (middle) and $\beta=0$ (right). Spatial correlation appears only for $\beta\neq 0$ (middle), while miscalibration-induced correlations are suppressed by ILC weights (right). \emph{Right block}: Regression in independent sky patches (top: $\beta \neq 0$; bottom: $\beta = 0$) yields a robust estimate of $\beta$.}
\label{fig:regression}
\end{figure}

\section{Results on \emph{LiteBIRD} simulations and \emph{Planck} PR4 data}

We validate the method on \emph{LiteBIRD} simulations (15 frequency channels spanning $40$--$402$\,GHz) including CMB, Galactic foregrounds, noise, polarization angle miscalibration $\alpha(\nu)$ across frequency channels, and either cosmic birefringence ($\beta = 0.3^\circ$) or not ($\beta = 0$). 

The birefringence angle $\beta$ is estimated as the regression coefficient between the two reconstructed fields, $(\tilde{E}^{\,\rm ILC} - \hat{E}^{\,\rm Hybrid\, ILC})(\hat{n})$ and $(F \ast \tilde{B}^{\,\rm ILC})(\hat{n})$. A strong spatial correlation between these fields is observed only when $\beta \neq 0$ (Figure~\ref{fig:regression}), while for $\beta = 0$ any correlation induced by miscalibration angles $\alpha(\nu)$ is suppressed due to spatial modulation by the ILC weights.

For $\beta = 0.3^\circ$, scatter plots between the two fields show consistent regression slopes across independent sky patches (Figure~\ref{fig:regression}), and linear regression yields $\hat{\beta} = 0.31^\circ \pm 0.02^\circ$, corresponding to a $>15\sigma$ detection.  In the absence of birefringence, no correlation is observed in the sky patches and $\hat{\beta}=0.01^\circ \pm 0.02^\circ$, which is consistent with zero, demonstrating that the method is unbiased.

We extracted CMB $E$- and $B$-mode maps from the \emph{Planck} PR4 data using both standard and Hybrid ILC methods, and computed the diagnostic observable $R_\ell$ (Eq.~\ref{eq:ratio}). We find $R_\ell$ to be relatively flat and constant across multipoles and sky fractions (Figure~\ref{fig:Rell}; right), in contrast with the strongly varying pattern induced by miscalibration in simulations with $\beta=0$ (Figure~\ref{fig:Rell}; left). The relative flatness of $R_\ell$ is strongly indicative of cosmic birefringence in the \emph{Planck} data.

Spatial regression between the reconstructed \emph{Planck} PR4 $(\tilde{E}^{\,\rm ILC} - \hat{E}^{\,\rm Hybrid\, ILC})$ and $(F \ast \tilde{B}^{\,\rm ILC})$ fields across independent sky patches yields an isotropic birefringence angle of $\hat{\beta} = 0.32^\circ \pm 0.12^\circ$, hence a $2.7\sigma$ detection, consistent with previous independent analyses \cite{pr4}. Moreover, our estimate remains stable across varying sky fractions, supporting a cosmological origin of the signal.

\section{Conclusions}

We have presented a blind, map-based ILC method to constrain cosmic birefringence, agnostic to foreground $EB$ correlations. We introduced the Hybrid $E$--$B$ ILC, which separates the correlated and uncorrelated components of the CMB polarization field and breaks the degeneracy between cosmic birefringence and polarization angle miscalibration by exploiting the achromatic nature of the former. The method enables a direct spatial regression estimator of the birefringence angle $\beta$. Applied to \emph{LiteBIRD} simulations, it yields constraints on $\beta$ that are competitive with and complementary to parametric power-spectrum approaches \cite{lb}. Applied to \emph{Planck} PR4 data, we find $\beta = 0.32^\circ \pm 0.12^\circ$, consistent with previous results \cite{pr4} and robust across sky fractions.

\section*{Acknowledgments}

My work is supported by the Spanish Ministry of Science and Innovation and the Agencia Estatal de Investigación through project grants PID2022-139223OB-C21 and PID2022-140670NA-I00.

\end{document}